\documentclass[preprint,superscriptaddress,nofootinbib]{revtex4}
     
     \usepackage{amssymb,amsmath,graphicx,subfigure,color}
     \usepackage{rotating,multirow,dcolumn}
     \usepackage[colorlinks,breaklinks]{hyperref}
     \allowdisplaybreaks[4]
     \begin{document}

     \title{Prospects for observing the ${\eta}_{t}$ ${\to}$ $W^{+}W^{-}$ decay at the HL-LHC}
     \author{Yueling Yang}
     \affiliation{School of Physics, Henan Normal University, Xinxiang 453007, China}
     \author{Bingbing Yang}
     \affiliation{School of Physics, Henan Normal University, Xinxiang 453007, China}
     \author{Junfeng Sun}
     \affiliation{School of Physics, Henan Normal University, Xinxiang 453007, China}

     \begin{abstract}
     Motivated by the recent observation of the pseudoscalar
     toponium ${\eta}_{t}$ at the LHC and by the
     ongoing interest in its properties,
     we phenomenologically investigate the ${\eta}_{t}$ ${\to}$
     $W^{+}W^{-}$ process, one of the characteristic
     decays of toponium ${\eta}_{t}$.
     The branching ratio for the ${\eta}_{t}$ ${\to}$ $W^{+}W^{-}$ decay
     is estimated with two scenarios of the decay constant
     $f_{{\eta}_{t}}$.
     It is found that tens of dilepton events from the cascade decays,
     ${\eta}_{t}$ ${\to}$ $W^{+}W^{-}$ ${\to}$
     ${\ell}^{-}\bar{\nu}_{\ell}{\ell}^{{\prime}+}{\nu}_{{\ell}^{\prime}}$
     with ${\ell}$, ${\ell}^{\prime}$ $=$ $e$ and ${\mu}$,
     could be experimentally observed, when considering the realistic
     identification efficiency of the charged lepton and
     assuming more than $3{\times}10^{7}$ ${\eta}_{t}$
     events available at the future HL-LHC.
     We propose searching for ${\eta}_{t}$ via the
     ${\eta}_{t}$ ${\to}$ $W^{+}W^{-}$ ${\to}$
     ${\ell}^{-}\bar{\nu}_{\ell}{\ell}^{{\prime}+}{\nu}_{{\ell}^{\prime}}$
     channel, which features no $b$-jets and clean dilepton signatures,
     thus offering a complementary and potentially more sensitive
     probe.

     \href{https://doi.org/10.1140/epjc/s10052-026-16112-1}{Eur. Phys. J. C 86, 882 (2026)}
     \end{abstract}
     \maketitle

     \section{introduction}
     \label{sec01}
     The top quark is the heaviest known fundamental fermion in
     the standard model (SM) of particle physics.
     It is generally believed that the lifetime of the top quark is
     too short to form stable topped bound states
     \cite{PLB.181.157,AnnRevNuclPartSci.53.301,
     JPhysG.35.083001,PhysRevD.110.030001}.
     Recently, however, a color-singlet $S$-wave
     $t\bar{t}$ bound state has been clearly observed
     by both the CMS and ATLAS groups using
     proton–proton collision data at a center-of-mass
     (CM) energy of $\sqrt{s}$ $=$ $13$ TeV at the CERN LHC
     \cite{RPP.88.127801,RPP.88.087801,RPP.89.057801}.
     This state is preferably identified as the ground spin-singlet
     pseudoscalar toponium ${\eta}_{t}$,
     with a mass $m_{{\eta}_{t}}$ ${\approx}$ $343$ GeV.
     The quest to understand the distinctive properties of the toponium
     ${\eta}_{t}$ will be a key focus of future research.

     The toponium ${\eta}_{t}$ is predominantly produced
     through the gluon–gluon fusion at the LHC colliders
     \cite{Z.Phys.C.48.613,PhysRevD.50.3173,PhysLettB.666.71,
     EPJC.60.375,JHEP.2010.09.034,PhysRevD.104.034023,
     PhysRevD.110.054032,PLB.866.139510,
     PLB.866.139532,EPJC.85.157,PhysRevD.112.094050,2506.14552}.
     Studies of toponium production at lepton
     colliders can be found in references of Ref. \cite{2506.14552}.
     At the CM energy of $\sqrt{s}$ $=$ $13$ TeV,
     the production cross section of toponium ${\eta}_{t}$
     has been
     measured to be ${\sigma}({\eta}_{t})$ $=$ $8.8^{+1.2}_{-1.4}$ pb
     by the CMS detector \cite{RPP.88.087801} and
     ${\sigma}({\eta}_{t})$ $=$ $9.3^{+1.4}_{-1.3}$ pb by
     the ATLAS detector \cite{RPP.89.057801}.
     It is expected that more than $3{\times}10^{7}$
     ${\eta}_{t}$ mesons could be produced with an integrated
     luminosity of $4$ ${\rm ab}^{-1}$  \cite{2505.03535}
     at the forthcoming HL-LHC experiment, which provides
     a solid foundation and a valuable opportunity to carefully
     investigate the ${\eta}_{t}$ meson.

     One of the most striking features of the ${\eta}_{t}$ meson
     is its decays.

     (1)
     Similar to the ground pseudoscalar charmonium ${\eta}_{c}$,
     which lies below the open-charm  threshold and thus
     cannot decay into two charmed hadrons,
     the ground pseudoscalar toponium ${\eta}_{t}$ is unlikely
     to decay into two top-flavored hadrons due to energy
     conservation.

     (2)
     In contrast to the ${\eta}_{c}$ meson, which decays dominantly
     via the annihilation of the $c\bar{c}$ pair into two gluons
     or two photons \cite{PLB.60.183,PhysRept.41.1,PhysRevD.37.3210},
     the toponium ${\eta}_{t}$ decays are overwhelmingly
     dominated by weak decays of its constituent top quarks.
     This is supported by the estimate
     ${\Gamma}_{{\eta}_{t}}$ ${\approx}$
     $2\,{\Gamma}_{t}$ ${\approx}$ $2.8$ GeV \cite{RPP.88.087801}
     and the measured top quark width
     ${\Gamma}_{t}$ ${\approx}$ $1.42^{+0.19}_{-0.15}$ GeV
     \cite{PhysRevD.110.030001}.

     (3)
     Both constituent quarks of the
     ${\eta}_{t}$ can decay individually.
     Given the hierarchy of the Cabibbo-Kobayashi-Maskawa (CKM)
     matrix elements, ${\vert} V_{tb} {\vert}$ ${\gg}$
     ${\vert} V_{ts} {\vert}$ ${\gg}$ ${\vert} V_{td} {\vert}$,
     the top quark decays almost exclusively into a bottom quark
     and an on-shell $W$ boson within the SM.
     Consequently, the toponium ${\eta}_{t}$ and
     nonresonant $t\,\bar{t}$ pair will share the same
     decay products of $W^{+}\,b\,W^{-}\,\bar{b}$ in most cases,
     making the separation of ${\eta}_{t}$ from the nonresonant
     $t\,\bar{t}$ backgrounds and nearby toponium states near
     the production threshold extremely challenging.

     (4)
     The large phase space available for
     the superheavy toponium ${\eta}_{t}$ decay can accommodate a
     rich variety of interesting
     final state topologies involving leptons and quarks.
     In addition to the dominant ${\eta}_{t}$ ${\to}$
     $W^{+}\,b\,W^{-}\,\bar{b}$ weak decay
     and the conventional decay modes ${\eta}_{t}$ ${\to}$ $gg$,
     ${\gamma}{\gamma}$, and $f\bar{f}$,
     the characteristic and potentially interesting channels include
     ${\eta}_{t}$ decays into two on-shell bosons, such as
     $W^{+}W^{-}$, $Z^{0}Z^{0}$ and $Z^{0}H$
     \cite{PhysRevD.35.3366,PhysRept.167.321,2506.14552,CPC.50.033101}\footnotemark[1].
     \footnotetext[1]{The ${\eta}_{t}$ ${\to}$ $HH$ decay
     is forbidden by Bose-Einstein
     statistics and $CP$ conservation
     \cite{PhysRevD.35.3366,PhysRept.167.321,2506.14552},
     considering the $J^{PC}$ $=$ $0^{-+}$ nature of
     the toponium ${\eta}_{t}$, and the fact that the
     two final scalar Higgs bosons are identical
     particles in an $S$-wave.}

     In this paper, we would like to reinvestigate the characteristic
     ${\eta}_{t}$ ${\to}$ $W^{+}W^{-}$ decay within the SM
     based on the following two considerations.

     (1)
     For the two-body ${\eta}_{t}$ ${\to}$ $W^{+}W^{-}$ decay,
     there are two less $b$-jets than the decay products of
     $W^{+}\,b\,W^{-}\,\bar{b}$ from the resonant and nonresonant
     $t\bar{t}$ decays.
     Experimentally, it is well known that the reconstruction of each
     track will bring certain errors for the final measurement results.
     The more particles in the final states,
     the lower the reconstruction efficiency and
     the greater the number of potential error sources.
     The absence of $b$-jets will surely eliminates the errors
     associated with $b$-jet identification and
     improves the signal reconstruction efficiency.

     (2)
     The estimated branching fractions exhibit a
     clear hierarchy, as shown below \cite{2506.14552}
     \begin{eqnarray}
    {\cal B}r({\eta}_{t}{\to}Z^{0}H) & {\sim} & 1.02{\times}10^{-3}
     \label{eta-zh-01}, \\
    {\cal B}r({\eta}_{t}{\to}W^{+}W^{-}) & {\sim} & 2.42{\times}10^{-4}
     \label{eta-ww-01}, \\
    {\cal B}r({\eta}_{t}{\to}Z^{0}Z^{0}) & {\sim} & 1.22{\times}10^{-5}
     \label{eta-zz-01},
     \end{eqnarray}
     or \cite{CPC.50.033101}
     \begin{eqnarray}
    {\cal B}r({\eta}_{t}{\to}Z^{0}H) & {\sim} & 1.79{\times}10^{-4}
     \label{eta-zh-02}, \\
    {\cal B}r({\eta}_{t}{\to}W^{+}W^{-}) & {\sim} & 3.25{\times}10^{-5}
     \label{eta-ww-02}, \\
    {\cal B}r({\eta}_{t}{\to}Z^{0}Z^{0}) & {\sim} & 3.11{\times}10^{-6}
     \label{eta-zz-02}.
     \end{eqnarray}
     The charged leptons in the $W^{-}$ ${\to}$ ${\ell}^{-}\bar{\nu}_{\ell}$
     and $Z^{0}$ ($H$) ${\to}$ ${\ell}^{+}{\ell}^{-}$ decays,
     particularly for ${\ell}$ $=$ $e$ and ${\mu}$,
     have well-definite energies and momenta
     in the rest frame of the parent bosons.
     This provides a clean signal that can be
     distinguished from the overwhelming hadronic
     backgrounds at the LHC experiments,
     while also ensuring a high identification efficiency.
     Considering the branching ratios
     \cite{PhysRevD.110.030001},
     \begin{eqnarray}
    {\cal B}r(W^{-}{\to}{\ell}^{-}\bar{\nu}_{\ell}) & {\sim} & 11\,\%
     \label{w-lv-01}, \\
    {\cal B}r(Z^{0}{\to}{\ell}^{+}{\ell}^{-}) & {\sim} & 3.4\,\%
     \label{z-ll-01}, \\
    {\cal B}r(H{\to}{\mu}^{+}{\mu}^{-}) & {\sim} & (2.6{\pm}1.3)\,{\times}\,10^{-4}
     \label{h-ll-01},
     \end{eqnarray}
     one can see that it would be easier to search
     for signals of the ${\eta}_{t}$ ${\to}$ $W^{+}W^{-}$ ${\to}$
     ${\ell}^{-}\bar{\nu}_{\ell}{\ell}^{{\prime}+}{\nu}_{{\ell}^{\prime}}$
     decay at the LHC experiments than those of the
     ${\eta}_{t}$ ${\to}$ $Z^{0}Z^{0}$, $Z^{0}H$ ${\to}$
     ${\ell}^{+}{\ell}^{-}{\ell}^{{\prime}+}{\ell}^{{\prime}-}$
     decays.
     Although the two neutrinos escape detection, the back-to-back
     topology of the two $W$  bosons in the rest ${\eta}_{t}$ frame
     can be exploited to constrain the missing momentum
     and reconstruct the $W^{+}W^{-}$ invariant mass.
     Additionally, the opposite-charge dileptons are helpful to
     recognize an unambiguous signature
     for the ${\eta}_{t}$ ${\to}$ $W^{+}W^{-}$ decay.

     \section{the ${\eta}_{t}$ ${\to}$ $W^{+}W^{-}$ decay within SM}
     \label{sec02}
     \begin{figure}[h]
     \includegraphics[width=0.5\textwidth,bb=200 530 360 635]{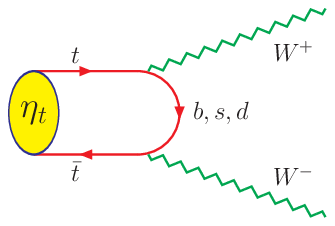}
     \caption{The lowest order Feynman diagram for the ${\eta}_{t}$
       ${\to}$ $W^{+}W^{-}$ decays within the SM.}
     \label{fig01}
     \end{figure}

     Due to the conservation of angular momentum, quantum number
     $J^{PC}$, and color charge, the ${\eta}_{t}$ decays via the
     quark-antiquark pair annihilation, ${\eta}_{t}$ ${\to}$ ${\gamma}^{\ast}$
     (or $g$, $Z^{0}$, $H$) ${\to}$ $W^{+}W^{-}$,
     are disallowed in principle,
     in spite of the large couplings of the
     $t\bar{t}H$ and $HW^{+}W^{-}$ vertices.
     Within the SM, the lowest order Feynman diagram for the ${\eta}_{t}$
     ${\to}$ $W^{+}W^{-}$ decay via quark exchange interaction
     \cite{PhysRevD.35.3366} is illustrated in Fig. \ref{fig01}.
     The $b$-quark-exchange amplitude can be written as,
     \begin{equation}
    {\cal A}_{b} \, = \,  {\rm Tr} \big\{
    {\langle} 0 {\vert} \bar{t}t {\vert} {\eta}_{t} {\rangle} \,
     \frac{ i\,g }{ 2\,\sqrt{2} }\, V_{tb}
     \not{\!\epsilon}^{\,\,{\ast}}_{W^{-}}\, (1-{\gamma}_{5}) \,
     \frac{i}{ \not{\!k}_{i}-m_{b} } \,
     \frac{ i\,g }{ 2\,\sqrt{2} }\, V_{tb}^{\ast}
     \not{\!\epsilon}^{\,\,{\ast}}_{W^{+}}\, (1-{\gamma}_{5}) \big\}
     \label{eq:amp-b},
     \end{equation}
     where the coefficient $g$ is the universal gauge coupling
     of the weak interaction, related to
     the Fermi constant $G_{F}$ by
     $g^{2}$ $=$ $\sqrt{32}\,m_{W}^{2}\,G_{F}$, with
     $G_{F}$ ${\approx}$ $ 1.166 \, {\times} \, 10^{-5} \, {\rm GeV}^{-2} $
     \cite{PhysRevD.110.030001}.
     $k_{i}$ denotes the momentum of the $b$ quark propagator.
     We will set the quark masses to zero,
     $m_{b}$ $=$ $m_{s}$ $=$ $m_{d}$ $=$ $0$,
     in the calculation, which
     is justified since that $m_{b}$ ${\ll}$ $m_{t}$
     and the exchanged quarks are off-shell.
     Using the unitarity of the CKM matrix,
     we have the relation,
     ${\vert} V_{tb} {\vert}^{2}$ $+$ ${\vert} V_{ts} {\vert}^{2}$ $+$
     ${\vert} V_{td} {\vert}^{2}$ $=$ $1$.
     The sum of the $b$-, $s$-, $d$-quark-exchange amplitudes
     is expressed as,
     \begin{eqnarray}
     i \, {\cal A}
     \, = \, \sum\limits_{i=d,s,b} i \, {\cal A}_{i}
     \, = \,  \frac{ G_{F} }{ \sqrt{2} }\,
     \frac{ m_{W}^{2} }{ k_{i}^{2} } \, {\rm Tr} \big\{
    {\langle} 0 {\vert} \bar{t}t {\vert} {\eta}_{t} {\rangle}
     \not{\!\epsilon}^{\,\,{\ast}}_{W^{-}}\, (1-{\gamma}_{5})
     \not{\!k}_{i}
     \not{\!\epsilon}^{\,\,{\ast}}_{W^{+}}\, (1-{\gamma}_{5}) \big\}
     \label{eq:amp-sum}.
     \end{eqnarray}

     Taking the notation of Ref. \cite{PLB.612.215,EPJC.60.107},
     the bilinear quark operator matrix elements associated with
     the pseudoscalar ${\eta}_{t}$ meson are decomposed as,
     \begin{equation}
    {\langle}\, 0\, {\vert}\, \bar{t}_{\alpha}(0)\, t_{\beta}(z)\,
    {\vert}\, {\eta}_{t}(p)\, {\rangle} \, = \,
     \frac{i}{4}\, f_{{\eta}_{t}}\, {\int}{\bf d}^{4}k\, {\rm e}^{+ik{\cdot}z}
     \big\{ \big[ \!\not{p}\,{\phi}_{{\eta}_{t}}^{a}(k)
     + m_{{\eta}_{t}}\,{\phi}_{{\eta}_{t}}^{p}(k) \big] \,
    {\gamma}_{5}  \big\}_{{\beta}{\alpha}}
     \label{eq:wf-etat},
     \end{equation}
     where $f_{{\eta}_{t}}$ is the decay constant.
     The scalar functions of ${\phi}_{{\eta}_{t}}^{a}$ and
     ${\phi}_{{\eta}_{t}}^{p}$ are respectively the twist-2
     and twist-3 wave functions.
     The calculation shows that the nonzero contribution to
     Eq.(\ref{eq:amp-sum}) arises only from the twist-2 component
     ${\phi}_{{\eta}_{t}}^{a}$.
     Both the wave function ${\phi}_{{\eta}_{t}}^{a}$ and
     decay constant $f_{{\eta}_{t}}$ are the undetermined
     hadronic parameters.

     The decay constant $f_{{\eta}_{t}}$ is an important
     observable that characterizes the toponium ${\eta}_{t}$.
     There are at least two possible scenarios for the
     decay constants of $S$-wave heavy quarkonium.
     One scenario (S1), based on the Royen-Weisskopf
     formula \cite{Nuovo.50.617},  assumes that the decay constant of
     nonrelativistic quark-antiquark bound systems depends on
     the (radial) wave function at the origin,
     \begin{equation}
     f_{{\eta}_{Q}} \, = \,
     \sqrt{ \frac{ 12 \, {\vert}{\psi}_{{\eta}_{Q}}(0) {\vert}^{2} }{ m_{{\eta}_{Q}} } }
     \, = \,
     \sqrt{ \frac{ 3 \, {\vert}R_{{\eta}_{Q}}(0) {\vert}^{2} }{ {\pi}\,m_{{\eta}_{Q}} } }
     \label{fetaQ-wf},
     \end{equation}
     where $R_{{\eta}_{Q}}(r)$ is radial wave function.

     According to the nonrelativistic QCD estimation
     \cite{PhysRevD.43.196,PhysRevD.46.4052,PhysRevD.51.1125,
     0412158,EPJC.71.1534},
     the typical squared valence-quark velocity\footnotemark[2]
     \footnotetext[2]{In the calculation, the quark masses
     are taken as \cite{PhysRevD.110.030001} $m_{c}$ $=$ $1.67(7)$ GeV,
     $m_{b}$ $=$ $4.78(6)$ GeV, and $m_{t}$ $=$ $172.57(29)$ GeV;
     the velocities are
     $v_{c}$ ${\sim}$ ${\alpha}_{s}(m_{c}v_{c})$ ${\approx}$ $0.572(14)$
     for charmonium,
     $v_{b}$ ${\sim}$ ${\alpha}_{s}(m_{b}v_{b})$ ${\approx}$ $0.343(2)$
     for bottomonium, and
     $v_{t}$ ${\sim}$ ${\alpha}_{s}(m_{t}v_{t})$ ${\approx}$ $0.147$
     for toponium, where the uncertainties come from the quark
     masses.}
     in the bound state are
     $v_{c}^{2}$ ${\sim}$ ${\alpha}^{2}_{s}$ ${\approx}$ $0.328(17)$
     for charmonium,
     $v_{b}^{2}$ ${\sim}$ ${\alpha}^{2}_{s}$ ${\approx}$ $0.118(1)$
     for bottomonium, and
     $v_{t}^{2}$ ${\sim}$ ${\alpha}^{2}_{s}$ ${\approx}$ $0.021$
     for toponium.
     The wave functions for heavy quarkonia can be obtained
     by solving the Schr\"{o}dinger equation with the interquark
     potentials consisting of short-range and long-range parts.
     The short-range potential, $V_{S}$,
     arises from the gluon-exchange
     interactions, while the long-range potential,
     $V_{L}$, is responsible for quark confinement.
     Phenomenologically, the Bohr radii,
     $r_{c}$ ${\sim}$ $1/m_{c}\,v_{c}$ ${\sim}$ $0.21$ fm
     for charmonium,
     $r_{b}$ ${\sim}$ $1/m_{b}\,v_{b}$ ${\sim}$ $0.12$ fm
     for bottomonium and
     $r_{t}$ ${\sim}$ $1/m_{t}\,v_{t}$ ${\sim}$ $0.01$ fm
     for toponium,
     lie deep within the short-range potential $V_{S}$ region and far
     away from the confining regime.
     The wave functions should be predominantly determined by
     $V_{S}$ with the spherically symmetric Coulomb-like
     form
     \cite{PhysRevLett.34.369,PhysLettB.66.286,PhysRept.56.167},
     \begin{equation}
     V(r) \, = \, - \, C_{F}\, \frac{{\alpha}_{s}}{r}
     \label{eq:Coulomb-potentials},
     \end{equation}
     where $C_{F}$ $=$ $4/3$, the QCD coupling constant at
     the energy scale ${\mu}$ is
     \begin{equation}
    {\alpha}_{s}({\mu}) \,  = \, 
     \frac{ 4\,{\pi} }
          { \displaystyle \Big( 11- \frac{2}{3}\,n_{f} \Big) \,
           {\ln} \Big( \frac{ {\mu}^{2} }{ {\Lambda}^{2} } \Big) }
     \label{qcd-as},
     \end{equation}
     and the specific value of QCD characteristic scale ${\Lambda}$
     varies with the flavor number $n_{f}$.

     The $1S$ wave function corresponding to the potential of
     Eq.(\ref{eq:Coulomb-potentials}) is given by,
     \begin{equation}
    {\psi}_{1S}(\vec{r}\,) \,  = \,
     \sqrt{ \frac{ q_{\rm B}^{3} }{ {\pi} } }\,
    {\exp}(-q_{\rm B}\,r)
     \label{eq:radial-wave-function},
     \end{equation}
     where $q_{\rm B}$ ${\approx}$ $C_{F}\,{\alpha}_{s}\,m_{Q}/2$
     is the Bohr momentum.
     By Fourier transformation of Eq.(\ref{eq:radial-wave-function})
     into the momentum-space representation ${\psi}_{1S}(\vec{k}\,)$,
     applying the momentum substitution ansatz
     \cite{PLB.612.215,EPJC.60.107,
      NPB.166.378,PhysRevD.32.1530,
      PhysRevD.37.778,PhysRevD.41.2319,PhysRevD.41.3394,
      PhysRevD.44.2851,PhysRevD.45.4214,PhysRevD.49.1490,
      PhysRevD.56.6010}: $\vec{k}_{z}$ ${\to}$ $(x-\bar{x})m_{0}/2$
      with $m_{0}^{2}$ $=$ $(m_{Q}^{2}+\vec{k}_{\perp}^{2})/x\,\bar{x}$,
      and integrating out the transverse momentum $\vec{k}_{\perp}$,
      the distribution amplitudes of the ${\eta}_{t}$ meson
      could be expressed as  \cite{PLB.612.215,EPJC.60.107},
     \begin{equation}
    {\phi}_{{\eta}_{t}}^{a}(x) \, = \,
     N\,  \frac{ (x\, \bar{x})^{2} }{ 1-4\,x\,\bar{x}\,(1-v_{t}^{2}) }
     \label{eq:wf-etat-v},
     \end{equation}
     where the variable $x$ denotes the longitudinal momentum fractions
     of the valence $t$ quark in the ${\eta}_{t}$ meson;
     $\bar{x}$ $=$ $1$ $-$ $x$ for the $\bar{t}$ quark;
     the normalization coefficient $N$
     is determined by
     \begin{math} 
    {\int}_{0}^{1} {\phi}_{{\eta}_{t}}^{a}(x) \,dx \, = \, 1
     \label{eq:wf-etat-normalization}.
     \end{math}
     Clearly, ${\phi}_{{\eta}_{t}}^{a}(x)$ is symmetric
     under the exchange $x$ ${\leftrightarrow}$ $\bar{x}$
     transformation. Consequently, terms proportional to
     ($x$ $-$ $\bar{x}$) in Eq.(\ref{eq:amp-sum})
     vanish and can be safely neglected.
     Finally, the partial decay width can be written as,
     \begin{eqnarray} & &
    {\Gamma}({\eta}_{t}{\to}W^{+}W^{-}) \, = \, 
     \frac{1}{8\,{\pi}} \, {\vert}\, {\cal A} \, {\vert}^{2} \,
     \frac{ {\vert} \vec{p}_{W} {\vert} }{ m_{{\eta}_{t}}^{2} }
     \nonumber \\ & = &
     \frac{ G_{F}^{2}\, f_{{\eta}_{t}}^{2} }{ 16\,{\pi} }\, m_{{\eta}_{t}}^{3}
     ( 1- 4\,z )^{3/2}
     \big( {\int} \frac{ {\phi}_{{\eta}_{t}}^{a}(x) }{ 1- x\,\bar{x}/z  }\,dx \big)^{2}
     \label{eq:width},
     \end{eqnarray}
     where ${\vert} \vec{p}_{W} {\vert}$ is the momentum of the $W$ boson
     in the rest frame of the ${\eta}_{t}$ meson,
     \begin{equation}
    {\vert} \vec{p}_{W} {\vert} \, = \,
     \frac{1}{2} \, m_{{\eta}_{t}}\, \sqrt{ 1- 4\,z }
     \label{eq:momentum-pw},
     \end{equation}
     and $z$ is the ratio of mass squared,
     $z$ $=$  $m_{W}^{2}/m_{{\eta}_{t}}^{2}$.

     With the Royen-Weisskopf formula of Eq.(\ref{fetaQ-wf})
     and wave function of Eq.(\ref{eq:radial-wave-function}),
     there is an approximate relation between the decay
     constant and the mass of the heavy quarkonium,
     \begin{equation}
     f_{{\eta}_{Q}} \ {\approx} \ m_{{\eta}_{Q}} \, {\alpha}_{s}^{3/2}
     \label{fetaQ-ma}.
     \end{equation}
     The decay constant of Eq.(\ref{fetaQ-ma}) will monotonically
     increase with the increase of mass of the heavy quarkonium
     when $m_{{\eta}_{Q}}$ $>$ $2$ GeV.
     Applying the relation of Eq.(\ref{fetaQ-ma})
     specifically to the case of
     ground charmonium ${\eta}_{c}$ and bottomonium ${\eta}_{b}$
     and toponium ${\eta}_{t}$, we obtain,
     \begin{eqnarray}
     f_{{\eta}_{c}} & {\approx} &
         m_{{\eta}_{c}} \, {\alpha}_{s}^{3/2}(m_{{\eta}_{c}})
         \, = \, 387.83(78) \, \text{MeV},
     \label{fetac-s1} \\
     f_{{\eta}_{b}} & {\approx} &
         m_{{\eta}_{b}} \, {\alpha}_{s}^{3/2}(m_{{\eta}_{b}})
         \, = \, 723.01(25) \, \text{MeV},
     \label{fetab-s1} \\
     f_{{\eta}_{t}} & {\approx} &
         m_{{\eta}_{t}} \, {\alpha}_{s}^{3/2}(m_{{\eta}_{t}})
         \, = \, 10.711(36) \, \text{GeV},
     \label{fetat-s1}
     \end{eqnarray}
     where the uncertainties come from the width effects
     $m_{{\eta}_{Q}}{\pm}{\Gamma}_{{\eta}_{Q}}/2$.
     \begin{table}[h]
     \caption{Decay constants (in the unit of MeV) of the
        ground pseudoscalar charmonium ${\eta}_{c}$ and
        bottomonium ${\eta}_{b}$.}
     \label{tab-fetac-fetab}
     \begin{ruledtabular}
     \begin{tabular}{l | rrr rrr}
     & $335(75)$\footnotemark[3]     \cite{PhysRevLett.86.30}   
     & $429(25)$\footnotemark[4]     \cite{PhysRevD.73.074507}  
     & $438(8)$\footnotemark[4]      \cite{PLB.651.171}         
     & $394.7(2.4)$\footnotemark[4]  \cite{PhysRevD.82.114504}  
     & $392.8(4.9)$\footnotemark[4]  \cite{PhysRevD.86.074503}  
     & $387(7)$\footnotemark[4]      \cite{NPB.883.306}         
       \\
     & $387(4)$\footnotemark[4]      \cite{EPJC.78.1018}        
     & $389(5)$\footnotemark[4]      \cite{EPJC.78.1018}        
     & $397.5(1.0)$\footnotemark[4]  \cite{PhysRevD.102.054511} 
     & $398.1(1.0)$\footnotemark[4]  \cite{PhysRevD.102.054511} 
     & $383.8(6.7)$\footnotemark[4]  \cite{CPC.48.123104}       
     & $394.4(3.4)$\footnotemark[4]  \cite{PhysRevD.111.054504} 
       \\
     & $320(40)$\footnotemark[5]  \cite{ZPC.57.43}             
     & $260(75)$\footnotemark[5]  \cite{JPG.39.015002}         
     & $309(39)$\footnotemark[5]  \cite{NPB.883.306}           
     & $453(4)$\footnotemark[5]   \cite{EPJC.75.45}            
     & $300(50)$                  \cite{PLB.339.270}           
     & $420(52)$                  \cite{ZPC.76.107}            
       \\
     & $424(25)$       \cite{ZPC.76.107}           
     & $420$           \cite{pramana.57.821}       
     & $292(25)$       \cite{PLB.596.84}           
     & $349$           \cite{PLB.615.79}           
     & $380$           \cite{PhysRevC.77.025203}   
     & $239$           \cite{PhysRevD.81.114019}   
       \\
     & $326$           \cite{PhysRevD.81.114019}   
     & $327$           \cite{PhysRevD.81.114019}   
     & $399$           \cite{PhysRevD.84.096014}   
     & $305$           \cite{PhysRevD.92.054031}   
     & $424.4$         \cite{IJMPE.25.1650059}     
     & $550$           \cite{PLB.753.330}          
       \\
      $f_{{\eta}_{c}}$
     & $371$           \cite{PLB.753.330}          
     & $359(10)$       \cite{EPJC.77.696}          
     & $343(9)$        \cite{EPJC.77.696}          
     & $385$           \cite{PLB.790.257}          
     & $424$           \cite{PhysRevD.74.014012}   
     & $402$           \cite{PhysRevD.74.014012}   
       \\
     & $326$                 \cite{PhysRevD.75.073016}  
     & $354$                 \cite{PhysRevD.75.073016}  
     & $593$                 \cite{PhysRevC.78.055202}  
     & $480$\footnotemark[6] \cite{PhysRevC.78.055202}  
     & $458$                 \cite{JPG.36.035003}       
     & $363$\footnotemark[6] \cite{JPG.36.035003}       
       \\
     & $471$                 \cite{JPG.38.085001}       
     & $360$\footnotemark[6]  \cite{JPG.38.085001}      
     & $231^{+52}_{-61}$      \cite{EPJC.73.2505}       
     & $304^{+115}_{-116}$    \cite{EPJC.73.2505}       
     & $353^{+22}_{-17}$      \cite{PhysRevC.92.055203} 
     & $543$                  \cite{AHEP.2018.5961031}  
       \\
     & $415$\footnotemark[6]  \cite{AHEP.2018.5961031}  
     & $578$                  \cite{AHEP.2018.5961031}  
     & $442$\footnotemark[6]  \cite{AHEP.2018.5961031}  
     & $350.3$                \cite{EPJC.78.592}        
     & $501$                  \cite{CPC.42.083101}      
     & $395$\footnotemark[6]  \cite{CPC.42.083101}      
       \\
     & $385^{+94}_{-93}$      \cite{JHEP.2020.12.065}   
     & $618$                 \cite{IJTP.59.2016}        
     & $516$\footnotemark[6] \cite{IJTP.59.2016}        
     & $375$                 \cite{IJTP.59.2016}        
     & $313$\footnotemark[6] \cite{IJTP.59.2016}        
     & $240(10)$             \cite{EPJC.82.889}         
       \\
     & $338(12)$             \cite{EPJC.82.1045}        
     & $356$                 \cite{PhysRevD.106.014009} 
     & $347$                 \cite{PhysRevD.106.014009} 
     & $403$                 \cite{PhysRevD.110.094021} 
     & $386$                 \cite{PhysRevD.110.094021} 
     & $349(6)$              \cite{PhysRevD.111.016011} 
       \\ \hline
     & $801(9)$\footnotemark[4]      \cite{PLB.651.171}         
     & $667(6)$\footnotemark[4]      \cite{PhysRevD.86.074503}  
     & $724(12)$\footnotemark[4]     \cite{PhysRevD.103.054512} 
     & $726.4(10.5)$\footnotemark[4] \cite{2603.01846}          
     & $500(100)$\footnotemark[5]    \cite{ZPC.57.43}           
     & $251(72)$\footnotemark[5]     \cite{JPG.39.015002}       
       \\
     & $811(34)$\footnotemark[5]  \cite{EPJC.75.45}             
     & $705(27)$                  \cite{ZPC.76.107}             
     & $709(20)$                  \cite{ZPC.76.107}             
     & $711$                      \cite{pramana.57.821}         
     & $287$                      \cite{PLB.615.79}             
     & $244$                      \cite{PhysRevD.81.114019}     
       \\
     & $414$                      \cite{PhysRevD.81.114019}     
     & $444$                      \cite{PhysRevD.81.114019}     
     & $708$         \cite{PhysRevD.84.096014}       
     & $1016.8$      \cite{IJMPE.25.1650059}         
     & $844$         \cite{PLB.753.330}              
     & $768$         \cite{PLB.753.330}              
       \\
     & $655(14)$     \cite{EPJC.77.696}              
     & $644(15)$     \cite{EPJC.77.696}              
     & $782$         \cite{FBS.59.133}             
     & $768$         \cite{FBS.59.133}             
     & $709$         \cite{PLB.790.257}            
     & $638$         \cite{PhysRevD.74.014012}     
       \\
     \multirow{2}{*}{$f_{{\eta}_{b}}$}
     & $599$                 \cite{PhysRevD.74.014012}  
     & $507$                 \cite{PhysRevD.75.073016}  
     & $897$                 \cite{PhysRevD.75.073016}  
     & $881$                 \cite{PhysRevC.78.055202}  
     & $755$\footnotemark[6] \cite{PhysRevC.78.055202}  
     & $867$                  \cite{JPG.36.035003}      
       \\
     & $744$\footnotemark[6]  \cite{JPG.36.035003}      
     & $834$                  \cite{JPG.38.085001}      
     & $694$\footnotemark[6]  \cite{JPG.38.085001}      
     & $605^{+32}_{-17}$      \cite{PhysRevC.92.055203} 
     & $665(90)$              \cite{PLB.758.118}        
     & $517$                  \cite{AHEP.2018.5961031}  
       \\
     & $431$\footnotemark[6]  \cite{AHEP.2018.5961031}  
     & $585$                  \cite{AHEP.2018.5961031}  
     & $488$\footnotemark[6]  \cite{AHEP.2018.5961031}  
     & $646$                  \cite{EPJC.78.592}        
     & $691^{+141}_{-~80}$    \cite{JHEP.2020.12.065}   
     & $940$                  \cite{IJTP.59.2016}       
       \\
     & $833$\footnotemark[6]  \cite{IJTP.59.2016}        
     & $671$                  \cite{IJTP.59.2016}        
     & $594$\footnotemark[6]  \cite{IJTP.59.2016}        
     & $410(10)$              \cite{EPJC.82.889}         
     & $647$                  \cite{PhysRevD.106.014009} 
     & $629$                  \cite{PhysRevD.106.014009} 
       \\
     & $646$                    \cite{PhysRevD.110.094021} 
     & $618$                    \cite{PhysRevD.110.094021} 
     & $654.8$                  \cite{EPJC.81.116}         
     & $578.2$\footnotemark[6]  \cite{EPJC.81.116}         
     & $639$                    \cite{EPJP.137.357}        
     & $529$\footnotemark[6]    \cite{EPJP.137.357}        
       \\
     & $590.1$                  \cite{PhysScr.99.095301}   
     & $550.2$\footnotemark[6]  \cite{PhysScr.99.095301} 
     & $630(9)$                 \cite{PhysRevD.111.016011}  
     \end{tabular}
     \end{ruledtabular}
       \begin{minipage}{0.13\textwidth}
       \footnotetext[3]{experiment.}
       \end{minipage}
       \begin{minipage}{0.14\textwidth}
       \footnotetext[4]{lattice QCD.}
       \end{minipage}
       \begin{minipage}{0.11\textwidth}
       \footnotetext[5]{sum rule.}
       \end{minipage}
       \begin{minipage}{0.42\textwidth}
       \footnotetext[6]{result including a QCD correction factor.}
       \end{minipage}
     \end{table}

     The decay constants $f_{{\eta}_{c}}$ and $f_{{\eta}_{b}}$
     obtained from experiment, lattice QCD, sum rule, and other
     theoretical methods are listed in Table \ref{tab-fetac-fetab}.
     It can be clearly seen from the numbers in Table. \ref{tab-fetac-fetab}
     that
     (1)
        There are still significant discrepancies among theoretical
        calculations of $f_{{\eta}_{c,b}}$ using different approaches.
     (2) The decay constants $f_{{\eta}_{c}}$ from both
         Eq.(\ref{fetac-s1}) and lattice QCD calculations
         \cite{PhysRevD.82.114504,PhysRevD.86.074503,
         NPB.883.306,EPJC.78.1018,PhysRevD.102.054511,CPC.48.123104,
         PhysRevD.111.054504} basically agree with the experimental
         measurement \cite{PhysRevLett.86.30},
         within experimental and theoretical uncertainties.
     (3) The decay constant $f_{{\eta}_{c}}$ in Eq.(\ref{fetac-s1})
         is very close to the lattice QCD results
         \cite{PhysRevD.82.114504,PhysRevD.86.074503,
         NPB.883.306,EPJC.78.1018,PhysRevD.102.054511,CPC.48.123104,
         PhysRevD.111.054504}.
         The decay constant $f_{{\eta}_{b}}$ in Eq.(\ref{fetab-s1})
         is consistent with recent lattice determinations within errors
         \cite{PhysRevD.103.054512,2603.01846}.
         This implies that the decay constant $f_{{\eta}_{t}}$
         in Eq.(\ref{fetat-s1}) might be regarded as
         an educated estimation.

     In addition to the S1 scenario based on
     Eq.(\ref{fetaQ-wf}) or Eq.(\ref{fetaQ-ma}),
     there is another possible scenario (S2) for
     the decay constants of pseudoscalar heavy quarkonia,
     which relies on a phenomenological
     scaling law \cite{NPB.406.340,PLB.362.173,
     PhysRevD.51.3613,PhysRevD.87.056001}
     that is independent of the heavy quark flavor, {\it i.e.},
     \begin{equation}
     \frac{ f_{{\eta}_{Q}}^{2} }{ m_{{\eta}_{Q}} } \ \approx \
     \text{const.}
     \label{fetaq-metaq}.
     \end{equation}
     This relation of Eq.(\ref{fetaq-metaq}) could be
     approximately verified using the experimental
     and lattice decay constants listed
     in Table \ref{tab-fetac-fetab}. For instance,
     \begin{equation}
     \frac{ f_{{\eta}_{c}} }{ \sqrt{ m_{{\eta}_{c}} } } \ \approx \
     \bigg[ \begin{array}{lll}
        0.194(43)  \text{ \cite{PhysRevLett.86.30} };
     &  0.227(3)   \text{ \cite{PhysRevD.86.074503}  };
     &  0.224(4)   \text{ \cite{NPB.883.306}         }; \\
        0.230(1)   \text{ \cite{PhysRevD.102.054511} };
     &  0.222(4)   \text{ \cite{CPC.48.123104}       };
     &  0.228(2)   \text{ \cite{PhysRevD.111.054504} };
     \end{array} \bigg] \, {\rm GeV}^{1/2},
     \label{fetac-metac-s2}
     \end{equation}
     \begin{equation}
     \frac{ f_{{\eta}_{b}} }{ \sqrt{ m_{{\eta}_{b}} } } \ \approx \
     \big[ \begin{array}{lll}
        0.218(2)  \text{ \cite{PhysRevD.86.074503}   };
     &  0.236(4)   \text{ \cite{PhysRevD.103.054512}  };
     &  0.237(3)   \text{ \cite{2603.01846}           };
     \end{array} \big] \, {\rm GeV}^{1/2}.
     \label{fetab-metab-s2}
     \end{equation}
     Additionally, for the ground vector charmonium $J/{\psi}$
     and bottomonium ${\Upsilon}(1S)$,
     the scaling relation of Eq.(\ref{fetaq-metaq})
     also seems to hold approximately\footnotemark[7].
     \footnotetext[7]{For a vector quarkonium with mass $m_{V}$,
     the relation between the decay constant $f_{V}$ and
     the experimental leptonic branching ratio is
     \cite{PhysRevD.92.074028,IJMPA.31.1650161},
     \begin{equation}
    {\cal B}r(V{\to}{\ell}^{+}{\ell}^{-}) \, = \,
     \frac{ 4{\pi}{\alpha}_{\rm em}^{2} }{ 3\,{\Gamma}_{V} }\,
     \frac{ Q_{q}^{2}\, f_{V}^{2} }{ m_{V} }\,
     \sqrt{\displaystyle 1-4\,\frac{ m_{\ell}^{2} }{ m_{V}^{2} } }
     \Big(\displaystyle 1+2\,\frac{ m_{\ell}^{2} }{ m_{V}^{2} }\Big)
     \label{fv-mv},
     \end{equation}
     where ${\alpha}_{\rm em}$ is the fine structure constant
     of electromagnetic interaction,
     ${\Gamma}_{V}$ is the decay width, and
     $Q_{q}$ is the electric charge number of valence quark,
     {\em i.e.}, $Q_{c}$ $=$ $+2/3$ for the charm quark
     and $Q_{b}$ $=$ $-1/3$ for the bottom quark.
     One can obtain $f_{J/{\psi}}$ $=$ $390.7(4.6)$ [$390.4(4.7)$] MeV
     with branching ratio
     ${\cal B}r(J/{\psi}{\to}e^{+}e^{-})$ $=$ $5.971(32)\,\%$
     [${\cal B}r(J/{\psi}{\to}{\mu}^{+}{\mu}^{-})$ $=$ $5.961(33)\,\%$]
     \cite{PhysRevD.110.030001} and the weighted average
     $f_{J/{\psi}}$ $=$ $390.6(3.3)$ MeV;
     $f_{{\Upsilon}(1S)}$ $=$ $658.8(18.6)$ [$671.1(13.2)$, $687.6(21.2)$] MeV
     with branching ratio
     ${\cal B}r({\Upsilon}(1S){\to}e^{+}e^{-})$ $=$ $2.39(8)\,\%$
     [${\cal B}r({\Upsilon}(1S){\to}{\mu}^{+}{\mu}^{-})$ $=$ $2.48(4)\,\%$,
     ${\cal B}r({\Upsilon}(1S){\to}{\tau}^{+}{\tau}^{-})$ $=$ $2.60(10)\,\%$]
     \cite{PhysRevD.110.030001} and the weighted average
     $f_{{\Upsilon}(1S)}$ $=$ $671.2(9.6)$ MeV.
     In the calculation,
     ${\alpha}_{\rm em}^{-1}$ ${\approx}$ $128.84$ [$128.60$] is
     used for the $J/{\psi}$ [${\Upsilon}(1S)$] decays.}
     \begin{equation}
     \frac{ f_{J/{\psi}} }{ \sqrt{ m_{J/{\psi}} } } \, = \,
     \frac{ 390.6(3.3)\,{\rm MeV} }{ \sqrt{ 3096.900(6)\,{\rm MeV} } } \, = \,
     0.222(2) \, {\rm GeV}^{1/2}
     \label{fpsi-mpsi-s2},
     \end{equation}
     \begin{equation}
     \frac{ f_{{\Upsilon}(1S)} }{ \sqrt{ m_{{\Upsilon}(1S)} } } \, = \,
     \frac{ 671.2(9.6)\,{\rm MeV} }{ \sqrt{ 9460.4(1)\,{\rm MeV} } } \, = \,
     0.218(3) \, {\rm GeV}^{1/2}
     \label{fy1s-my1s-s2}.
     \end{equation}

     Based on the above cases of charmonium and bottomonium
     under the S2 scenario,
     we assume that the scaling law in
     Eq.(\ref{fetaq-metaq}) also applies to the
     case of toponium ${\eta}_{t}$.
     As a conservative estimate, we take
     \begin{equation}
     \frac{ f_{{\eta}_{t}} }{ \sqrt{ m_{{\eta}_{t}} } } \ \approx \
     0.22(5) \, \, {\rm GeV}^{1/2}
     \label{eq:ratio-feta-meta}.
     \end{equation}
     This yields a decay constant
     $f_{{\eta}_{t}}$ ${\approx}$ $4.07(93)$ GeV,
     which is less than half the value obtained from the S1 scheme in
     Eq.(\ref{fetat-s1}).

     Using Eq.(\ref{eq:width}), the partial decay width, branching
     ratio, and event numbers are respectively estimated to be,
     \begin{eqnarray}
    {\Gamma}({\eta}_{t}{\to}W^{+}W^{-}) & {\approx} & 807.7(67.2)\, {\rm keV}
     \label{width-s1}, \\
    {\cal B}r({\eta}_{t}{\to}W^{+}W^{-}) & {\approx} & 28.8(2.4) {\times} 10^{-5}
     \label{br-s1}, \\
    N_{{\eta}_{t}{\to}W^{+}W^{-}} & {\approx} & 8654(720)
     \label{nww-s1},
     \end{eqnarray}
     for the S1 scenario, and
     \begin{eqnarray}
    {\Gamma}({\eta}_{t}{\to}W^{+}W^{-}) & {\approx} & 116.9(70.2)\, {\rm keV}
     \label{width-s2}, \\
    {\cal B}r({\eta}_{t}{\to}W^{+}W^{-}) & {\approx} & 4.2(2.5) {\times} 10^{-5}
     \label{br-s2}, \\
    N_{{\eta}_{t}{\to}W^{+}W^{-}} & {\approx} & 1252(752)
     \label{nww-s2},
     \end{eqnarray}
     for the S2 scenario, where the overall uncertainties of
     branching ratio come from the
     decay constant $f_{{\eta}_{t}}$, the width effects of
     $m_{{\eta}_{t}}{\pm}{\Gamma}_{{\eta}_{t}}/2$ and
     $m_{W}{\pm}{\Gamma}_{W}/2$, with
     the width ${\Gamma}_{{\eta}_{t}}$ $=$ $2.8$ GeV \cite{RPP.88.087801}.
     The branching ratio of the S1 (S2) scenario is
     comparable to
     the estimation of Ref. \cite{2506.14552} (Ref. \cite{CPC.50.033101}).
     For the event number estimates, we assume a
     prospective data sample of $3{\times}10^{7}$ ${\eta}_{t}$
     events at the future HL-LHC.
     The large discrepancy between the S1 and S2 scenarios highlights
     the theoretical uncertainty in predicting the decay constant
     $f_{{\eta}_{t}}$.
     In the Royen-Weisskopf S1 scenario, the decay constant
     and wave function are closely related to the interquark
     potentials and the understanding of the interaction mechanism
     among constituents of heavy quarkonium.
     The approximation $f_{{\eta}_{Q}}$ in Eq.(\ref{fetaQ-ma})
     is obtained with only the Coulomb-like potential,
     although the $f_{{\eta}_{c}}$ in Eq.(\ref{fetac-s1})
     and $f_{{\eta}_{b}}$ in Eq.(\ref{fetab-s1})
     agree with the lattice calculation within uncertainties.
     However, when a more rigorous and complete interquark potential
     that includes confining and spin-dependent hyperfine interactions
     is employed, large uncertainties persist, as reflected in
     the results \cite{PhysRevD.74.014012,
     PhysRevD.75.073016,PhysRevC.78.055202,JPG.36.035003,
     JPG.38.085001,AHEP.2018.5961031,EPJC.78.592,
     CPC.42.083101,JHEP.2020.12.065}
     listed in Table \ref{tab-fetac-fetab}.
     The scaling law in Eq.(\ref{fetaq-metaq}) of S2 scenario is
     more of an empirical generalization, whose underlying
     dynamics require further investigation.
     Lattice calculations might provide more reliable results;
     however, the large top quark mass demands an exceedingly
     small lattice spacing, so much work still lies ahead.
     Experimental measurements of the decay constant $f_{{\eta}_{Q}}$
     are therefore highly desirable.

     We would like to propose the purely leptonic two-body decays
     of the $W$ bosons to make the searching for the ${\eta}_{t}$
     ${\to}$ $W^{+}W^{-}$ decay easier to stand out from the
     chaotic hadron background.
     The typical selection efficiency ${\epsilon}_{\ell}$ of
     well-identified electrons (muons) with ``tight'' identification
     criteria is about $70\,\%$ ($75\,\%$-$85\,\%$)
     \cite{RPP.88.127801,JINST.13.P06015,JINST.16.P05014,
     JINST.14.P12006,EPJC.81.578,EPJC.83.686,EPJC.86.470}
     at the CMS and ATLAS detectors, which may vary depending on
     the exact trigger and reconstruction strategy.
     Considering the charged lepton production rate from the $W$
     boson decay \cite{PhysRevD.110.030001}, Eq.(\ref{w-lv-01}),
     the number of observable events from the
     cascade decays, ${\eta}_{t}$ ${\to}$ $W^{+}W^{-}$,
     $W^{+}$ ${\to}$ ${\ell}^{+}{\nu}_{\ell}$,
     $W^{-}$ ${\to}$ ${\ell}^{{\prime}-}\bar{\nu}_{{\ell}^{\prime}}$,
     could be estimated with the following formula,
     \begin{equation}
     N_{\rm obs} \, = \, N_{{\eta}_{t}{\to}W^{+}W^{-}} \, {\times} \,
    {\cal B}r(W^{+}{\to}{\ell}^{+}{\nu}_{\ell}) \, {\times} \,
    {\epsilon}_{\ell} \, {\times} \,
    {\cal B}r(W^{-}{\to}{\ell}^{{\prime}-}\bar{\nu}_{{\ell}^{\prime}}) \,
    {\times} \, {\epsilon}_{{\ell}^{\prime}}
     \label{n-cascade}.
     \end{equation}
     The results listed in Table \ref{tab-num-cascade}
     demonstrate a realistic and promising possibility
     for investigating the cascade decays,
     ${\eta}_{t}$ ${\to}$ $W^{+}W^{-}$ ${\to}$
     ${\ell}^{+}{\nu}_{\ell}{\ell}^{{\prime}-}\bar{\nu}_{{\ell}^{\prime}}$,
     at the future HL-LHC.

     \begin{table}[t]
     \caption{The possible event numbers of the experimentally observable
     ${\eta}_{t}$ ${\to}$ $W^{+}W^{-}$ ${\to}$
     ${\ell}^{+}{\nu}_{\ell}{\ell}^{{\prime}-}\bar{\nu}_{{\ell}^{\prime}}$
     decays, where the electron (muon) selection efficiency
     is set approximately to $70\,\%$ ($80\,\%$).}
     \label{tab-num-cascade}
     \begin{ruledtabular}
     \begin{tabular}{c ccc}
      final state & $e^{+}e^{-}{\nu}_{e}\bar{\nu}_{e}$
                  & $e^{+}{\mu}^{-}{\nu}_{e}\bar{\nu}_{\mu}$ $+$ $c.c$
                  & ${\mu}^{+}{\mu}^{-}{\nu}_{\mu}\bar{\nu}_{\mu}$ \\ \hline
      S1 scenario & $49(4)$ & $110(9)$ & $63(5)$ \\
      S2 scenario & $ 7(4)$ & $16(10)$ & $ 9(5)$
     \end{tabular}
     \end{ruledtabular}
     \end{table}

     It should be noted here that the actual observability of
     the cascade decays, ${\eta}_{t}$ ${\to}$ $W^{+}W^{-}$ ${\to}$
     ${\ell}^{+}{\nu}_{\ell}{\ell}^{{\prime}-}\bar{\nu}_{{\ell}^{\prime}}$,
     depends critically on the relevant backgrounds, especially
     the irreducible process $pp$ ${\to}$ $W^{+}W^{-}$ ${\to}$
     ${\ell}^{+}{\nu}_{\ell}{\ell}^{{\prime}-}\bar{\nu}_{{\ell}^{\prime}}$.
     Although the production cross section for the background $pp$ ${\to}$
     $W^{+}W^{-}$ process falls rapidly above the $W^{+}W^{-}$
     threshold \cite{JHEP.2508.142,PhysRevD.102.092001},
     it might still reach a few fb in the ${\eta}_{t}$ mass window,
     which would be comparable to, or even larger than, the signal
     cross section for ${\sigma}({\eta}_{t})
    {\times} {\cal B}r ({\eta}_{t} {\to} W^{+}W^{-})$.
     Kinematic selections on the diboson $WW$ system from the
     signal ${\eta}_{t}$ topology, along with the cut 
     ${\vert}\vec{p}_{W}{\vert}$ ${\in}$ $(150,153)$ GeV in the
     ${\eta}_{t}$ rest frame [see Eq.(\ref{eq:momentum-pw})],
     could be exploited to reject the bulk of the threshold and
     continuum backgrounds from diboson $W^{+}W^{-}$ production.
     Nevertheless, the residual $pp$ ${\to}$ $W^{+}W^{-}$ background
     remains and must be further suppressed.
     Therefore, developing a more efficient method to suppress
     the background is crucial for the
     successful reconstruction of the ${\eta}_{t}$ ${\to}$ $W^{+}W^{-}$
     signal, and deserves dedicated in-depth investigation.
     Effectively reducing the background and enhancing the signal
     identification efficiency demand not only more expertise
     and technical proficiency but also thorough and meticulous
     studies, which unfortunately lie beyond the scope of
     our current capabilities.
     It is expected that, after imposing all selection requirements
     and employing a robust experimental reconstruction algorithm,
     an event excess near the ${\eta}_{t}$ kinematic threshold might
     be observed above the nonresonant $W^{+}W^{-}$ continuum,
     similar to what has been observed for the ${\eta}_{t}$
     resonance in $t\bar{t}$ production.

     \section{summary}
     \label{sec03}
     The recent observation of a resonance-like pseudoscalar toponium
     ${\eta}_{t}$ by both the CMS and ATLAS experimental groups
     at the LHC has renewed interest in the properties of
     toponium, although some controversies remain.
     More sufficient and precise measurements are urgently needed.
     The ${\eta}_{t}$ ${\to}$ $W^{+}W^{-}$ decay is one of the
     characteristic and distinctive processes to identify the
     toponium ${\eta}_{t}$.
     Encouraged by the optimistic prospect of more
     than $3{\times}10^{7}$ ${\eta}_{t}$ events at the
     future HL-LHC, we investigated the ${\eta}_{t}$
     ${\to}$ $W^{+}W^{-}$ decay within the SM.
     The decay constant of toponium is estimated
     to be either $f_{{\eta}_{t}}$ $=$ $10.711(36)$ GeV in the
     Royen-Weisskopf (S1) scenario or $f_{{\eta}_{t}}$ $=$
     $4.07(93)$ GeV in the scaling law (S2) scenario.
     The resulting branching ratios in the S1 and S2 scenario agree
     basically with the previous estimates of Ref. \cite{2506.14552}
     and \cite{CPC.50.033101}, respectively.
     The difference in the values of $f_{{\eta}_{t}}$
     gives rise to significant discrepancies in
     branching ratios between the
     S1 and S2 scenarios, which indicates our understanding
     of toponium's properties is still far from comprehensive.
     For the cascade decay, ${\eta}_{t}$ ${\to}$ $W^{+}W^{-}$ ${\to}$
     ${\ell}^{-}{\ell}^{{\prime}+}\bar{\nu}_{\ell}{\nu}_{{\ell}^{\prime}}$
     with ${\ell}$, ${\ell}^{\prime}$ $=$ $e$, ${\mu}$,
     the diboson $WW$ system with an invariant mass $m_{WW}$ within the
     ${\eta}_{t}$ mass window, together with
     the positively and negatively charged leptons with a definite
     momentum would provide helpful signal selection criteria.
     Considering the realistic identification efficiency of the
     charged lepton with the unique experimental environment,
     the number of observable signal events are predicted.
     We conclude that it might be feasible,
     albeit very challenging,
     to experimentally explore and investigate
     the ${\eta}_{t}$ ${\to}$ $W^{+}W^{-}$ decay
     at the future HL-LHC.
     This study aims to provide a useful reference for future
     experimental searches for toponium decays at the HL-LHC.

     \section*{Acknowledgments}
     The work is supported by the National Natural Science Foundation of China (Grant No. 12275068),
     National Key R\&D Program of China (Grant No. 2023YFA1606000),
     Natural Science Foundation of Henan Province (Grant Nos. 262300421308, 252300421491, 242300420250).

     

     \end{document}